\documentclass[twoside,reqno,10pt]{article}

\usepackage{amsmath}
\usepackage{graphicx,color,psfrag, amssymb}
  \usepackage{epsfig}

	\newcommand{\frdiffi}[3]{\ensuremath{\,_{#3}\mathbf{D}^{#1}_#2}}
	
	\newcommand{\frdiffix}[2]{\frdiffi{#1}{ #2}{ }}
	
	\newcommand{\frdiffii}[3]{\ensuremath{\,_{#3}\mathbf{I}^{#1}_#2}}
	
	\newcommand{\frdiffiix}[2]{\frdiffii{#1}{x}{#2}}
	
	\newcommand{\frdiffir}[3]{\ensuremath{\,_{#3 RL}\mathbf{D}^{#1}_#2}}
	
	\newcommand{\frdiffixr}[2]{\frdiffir{#1}{#2}{ }}

	\newcommand{\epnt}{\; .}
	\newcommand{\ecma}{\; ,}

\begin{document}
  \title{A model of fractional-order diffusion in the glial scar}
   
   \author {Dimiter Prodanov\footnote{Correspondence: Environment, Health and Safety, IMEC vzw, Kapeldreef 75, 3001 Leuven, Belgium;  
   		e-mail: Dimiter.Prodanov@imec.be, dimiterpp@gmail.com } \ and Jean Delbeke}


\maketitle

\begin{abstract}
	
	Implantation of neuroprosthetic electrodes induces a stereotypical state of neuroinflammation, which is thought to be detrimental for the neurons surrounding the electrode.
	Mechanisms of this type of neuroinflammation are still not understood well. 
	Recent experimental and theoretical results point out possible role of the diffusion species in this process.   
	 
	The paper considers a model of  anomalous diffusion occurring in the glial scar around a chronic implant in two simple   geometries -- a separable  rectilinear electrode and a cylindrical electrode, which are solvable exactly.
	We describe a hypothetical extended source of diffusing species and study its concentration profile in steady-state conditions.
	Diffusion transport is assumed to obey a fractional-order Fick law, which is derived from physically realistic assumptions using a fractional calculus approach. 
    The derived fractional-order distribution morphs into regular order diffusion in the case of integer fractional exponents. 
	The	model presented here demonstrates that accumulation of diffusing species can occur and the scar properties (i.e. tortuosity,  fractional order, scar thickness) can influence such accumulation.
	The observed shape of the concentration profile corresponds qualitatively with GFAP profiles reported in the literature. 
	The main difference with respect to the previous studies is the explicit incorporation of the apparatus of fractional calculus without assumption of an ad hoc tortuosity parameter.	
	Intended application of the approach is the study of diffusing substances in the glial scar after implantation of neural prostheses, although the approach can be adapted to other studies of diffusion in biological tissues, for example of biomolecules or small drug molecules. 

\end{abstract}

\section{Introduction}
\label{sec:intro}

Implantation of neuroprosthetic electrodes induces a sustained state of neuroinflammation and scarring, which is thought to be detrimental for the neurons surrounding the electrode \cite{McConnell2009a}.
Literature demonstrates that in chronic conditions the recording longevity of such electrodes in experimental animals is highly variable (i.e. for wire electrodes - \cite{Liu1999}, silicon-based electrodes and multi-wire arrays  --\cite{Ward2009}).
Over 100 studies have described stereotypic features of the brain response to microelectrodes that occur irrespective of the type of implant, method of sterilization, species studied,
or implantation method \cite{Jorfi2015}.

\begin{figure}[!ht]
	\begin{center}	
 \includegraphics[width=1.0\linewidth, bb=0 0 1421 688]{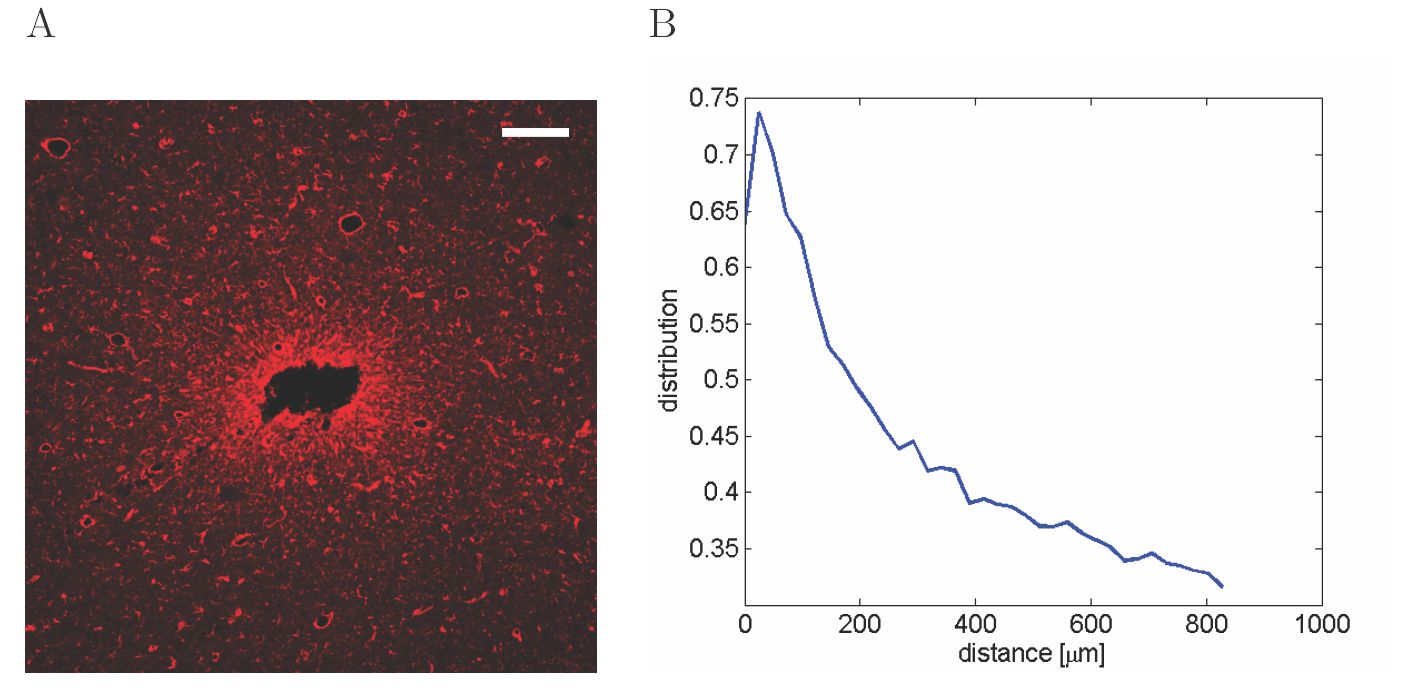}
\caption{GFAP distribution around an implant site}
\label{fig:gfapdist}
\end{center}
A -- GFAP staining after 6 weeks of implantation in tethered configuration; B -- Mean intensity distribution as a function of the distance to the insertion track \cite{Prodanov2012}.  
The dataset was published previously in \cite{Welkenhuysen2011}. Scale bar -- 200 um. 
\end{figure}

The formation of the glial scar is a complex reactive process involving  interactions between several types of cells, notably astrocytes and activated microglia, which are mediated by plethora of bio-active molecules (i.e. cytokins).  
The reactive \textit{astrocytes} form a dense web of interdigitated processes which over-expresses the Glial Fibrillary Acidic Protein (GFAP)  and attach to the implant (see Fig. \ref{fig:gfapdist}). 
Thus GFAP  is commonly used in neuroprosthetic studies as a marker of neuroniflammation.
In parallel, the \textit{microglia} are rapidly activated in a wide area around the lesion site and undergoe a profound change in cell shape and phenotype. 
Concurrent with the glial scar formation, neuronal density within the recording
radius of the microelectrodes is reduced, leading to even fewer
distinguishable single-unit recordings \cite{Edell1992,Turner1999,Kim2004,Biran2005, Purcell2009}.  
During this process the cells change substantially the composition, the morphology and the functional properties of the extracellular matrix. 
Roitbak and Sykov\'{a} \cite{Roitbak1999} also demonstrate changes of the diffusion path in reactive astrogliosis states and therefore in the extracellular space (ECS) properties. 

Several hypotheses for these observations have been put forward. 
Some authors proposed that formation of \textit{glial scar results in a physical barrier} to diffusing substances, thus creating a toxic environment for the neurons.
Others have proposed that many neurons around the electrodes die shortly after implantation \cite{Edell1992, Biran2005}. 
More recently, McConnell et al. \cite{McConnell2009a} have proposed that the observed loss of signal  can also result from the \textit{progressive degeneration of nerve fibers and synapses} due to persistent local chronic inflammation. 
Ward et al. attribute electrode failure to the traumatic  injury  resulting from insertion and a long-term foreign body response to the implant \cite{Ward2009}.

While there are observations supporting each hypothesis there is no agreement as to which is the predominant effect at different time scales (review in \cite{Polikov2005,Leach2010}). 
Quantitative histological descriptions of the glial scar may therefore demonstrate these effects more convincingly and provide metrics for more robust safety and biocompatibility assays for neural prostheses.

In this paper we study accumulation of a diffusing species in steady-state   conditions, produced  as a reaction to the presence of another object, i.e. an implanted electrode. 
We hypothesize that such conceptual model can describe diffusion phenomena occurring in the extracellular matrix following  implantation of flat electrodes, for example Michigan type of silicon probes or cylindrical electrodes, such as microwires.

\section{Extended source compartment model}	
\label{sec:expdiff}
 
Implanted electrodes have simple cross-section geometries and high aspect ratio with regard to height.
In such way, implant geometry can be approximated as consisting of ideal shapes, i.e. an infinite cylinder, a line or a half-plane. 
This allows for applying a symbolic calculation approach in the modeling of the diffusion problem.

\begin{figure}[h!]
	\centering
	\includegraphics[width=0.5\linewidth, bb=0 0 825 254]{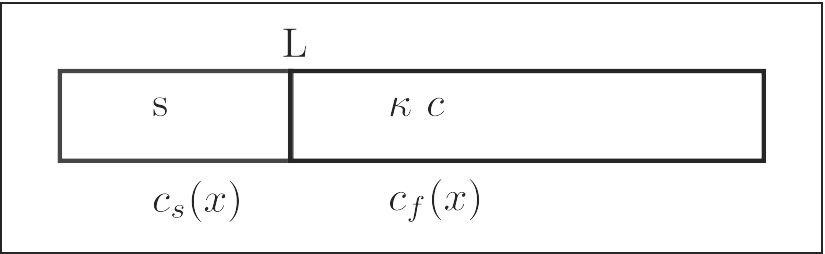}
	\caption{Model geometry: The extended source model comprises two compartments: source  (S) and tissue  (T).}
	\label{fig:model1}
\end{figure}

As an idealized situation we will consider semi-infinite medium with \textbf{two} spatial compartments: 
a \textsl{source compartment} (S) and a \textsl{tissue compartment} (T), (see Fig. \ref{fig:model1}). 
To simplify calculation we will assume that the S-compartment will have only constant production of  diffusing species but no degradation, while the T-compartment will have only first order degradation but no production.

The overall solution will be represented by a sum of two terms
\[
c (x)   = c_s(x) + c_f(x) 
\]
for the two coupled domains.
In the following presentation 
the  term $c_s$ will represent the concentration of the substance in
in the glial scar, i.e. the S-compartment. 
while the second term $c_f$ will represent the concentration in the outer zone -- i.e. the  T-compartment.

Further, the terms are assumed to be orthogonal, i.e. $c_s(x)=0$ in the distal compartment $x>L$  whereas  $c_f(x)=0$ in the proximal compartment $x<L$.
Since the overall solution is assumed to be continuous the following conditions will have to be  imposed on the boundary between domains:

\begin{flalign*} 
\left.  c_s(x) \right|_{x = L-} &= \left.  c_f(x) \right|_{x= L+} \\
\left. \frac{\partial c_s}{ \partial x} (x)\right|_{x = L-} &= \left. \frac{\partial c_f}{ \partial x} (x) \right|_{x = L+}
\end{flalign*}
In addition, we also assume  natural boundary conditions in steady state -- 
$ 	c (0) = c_0 , \ \ 	c(\infty)  =0 $.
Subsequent analysis is performed both in traditional partial differential equation setting and in fractional calculus (i.e. differ-integral) approach as a possible conceptual generalization.  

\section{Anomalous diffusion in complex media}
\label{sec:secdif}

At present it is well established that the ECS occupies a volume fraction of between 15 and 30\% in normal adult brain tissue with a typical value of 20\% and that this falls to 5\% during global ischemia \cite{Sykova2008}.
Therefore, the diffusion impediment can not be neglected in modeling of biological diffusion. 
Impeded diffusion can be modeled by two classes of models. 

\subsection{Regular impeded diffusion}
\label{sec:regdiff}

In this framework, the medium imposes only a spatial impediment on the diffusing species and does not change the diffusion law. 
Nicholson et al \cite{Nicholson2000, Nicholson2001} consider that the densely packed cells of the brain and their interstitial spaces resemble a porous medium with two phases, an intra- and extracellular phase.
Diffusion in the permeable phase of porous media is analogous to the diffusion in the narrow spaces between brain cells, i.e. the extracellular space, ECS.
 
In accordance, volume transport of species having concentration \textit{c} is governed by the rescaled 
reaction-diffusion equation:
\[
\frac{\partial c}{ \partial t}  = \frac{D}{ \lambda^2}   \nabla^2 c  +\frac{s}{\alpha} - \kappa c , \ \ \lambda=\sqrt{\frac{D}{\bar{D}}}
\]
where  the source term is denoted by \textit{s}  (1/concentration) , the clearance term denoted by $k (c) =\kappa \, c $ is assumed to be first order decay and \textit{D} is the diffusion coefficient (length$^2$/ time).
The following additional parameters are represented respectively by the parameters
$\lambda$ -- tortuosity (dimensionless);
$\alpha$ -- porosity coefficient (dimensionless);
$\kappa$ --  clearance speed ( 1/time).


Unless stated explicitly, we will further use a re-parametrization of the problem preserving the form and  interpretation of the equations: 
\[
 \ \bar{D} =\dfrac{D}{\lambda^2} , \ \ \bar{s} =\frac{s}{\alpha}
\]

	

\subsection{Fractional diffusion phenomena}
\label{sec:frdiff1}

Since the brain extracellular space is a complex medium this transport equation can be regarded only as an \textit{linear approximation}. 
In accordance,  tortuosity can be considered as the linear correction for the anomalous diffusion \cite{Nicholson2001} and \cite{Sykova2008}.

Diffusion in porous media, such as tissues, are characterized by deviation from the usual Fick's diffusion laws.
Notably, these processes are characterized by a distribution of waiting times, having
heavy tails approximated by an inverse fractional power law, and hence, dependence of the effective diffusion phenomenon on the duration of the measurement. 
Also the law of the mean squared displacements exhibits deviations from linearity.  
The resulting behavior can not be described by an ordinary differential equation.  

Since the end of the XX\textsuperscript{th} century a different approach to handle deviations from the classical diffusion and hence the anomalous diffusion phenomena has been employed. 
In this framework the underlying physical process is modeled  as a \textit{continuous time random walks} \cite{Metzler2004}, using the mathematical apparatus  of fractional calculus \cite{Oldham1974}.
In the fractional calculus approach the transport equation reads
 \[
 \frac{\partial c}{ \partial t}  = D  \;  \nabla \cdot \nabla^\beta  c + s- \kappa c 
 \]
where $\nabla^\beta$ denotes the fractional-order Fick's law, i.e. the fractional flux in the system.
Interpretation for the symbol of the last operator will be given further in the relevant sections of the paper.
This form of the fractional Fick's law naturally implies spatial non-locality and can be derived from rigorous approaches using spatial averaging theorems \cite{Meerschaert2006, Wheatcraft2008}.

	\section{Linear geometry}
	\label{sec:linear}
	
	\subsection{Regular diffusion along the line}
	\label{sec:diff1d}
	
	In this case we consider the following system:
		\begin{flalign*}
 		\frac{\partial c_s}{ \partial t} &=	D   \frac{\partial^2 c_s}{ \partial x^2}   - \kappa \, c_s  \\
		\frac{\partial c_f}{ \partial t} &=	D   \frac{\partial^2 c_f}{ \partial x^2}  + s    \\
		\end{flalign*}	 		
	In steady state the equations have the following general solutions:
	\begin{flalign*} 
		c_s(x) &=-\frac{s\,{x}^{2}}{2\,D}+k_5\,x+k_3 \\
	    c_f(x) &=k_1\,{e}^{\sqrt{\frac{\kappa}{D}}\, x}+k_2\,{e}^{-\sqrt{\frac{\kappa}{D}}\, x}
	\end{flalign*}
	Then since the solution is limited $k_1=0$. 
	After some algebraic transformations we arrive at the following algebraic system to be solved
	\begin{flalign*} 
	-\frac{s\,{L}^{2}}{2\,D}+k_5\,L+k_3 & =k_2\,{e}^{-\sqrt{\frac{\kappa}{D}}\, L}  \\
	k_5-\frac{s\,L}{D} & =-\frac{k_2\,\sqrt{\kappa}\,{e}^{-\sqrt{\frac{\kappa}{D}}\, L}}{\sqrt{D}}
	\end{flalign*}
	Assuming further parametrization by $k_3=c(0)= c_0$ we get
	
	\begin{flalign*} 
	c_s(x) &=-\frac{s}{2\,D} \,x^{2}+ 
	\frac{  \sqrt{\kappa} s L^2 + 2s \sqrt{D} L -2 \, c_0 \sqrt{\kappa}D  }{ 2\,D\,\left( \sqrt{\kappa} \,L+ \, \sqrt{D}  \right) } \, x
	+c_0 , \ \ x \in [0, L]\\
	c_f(x) & =  \frac{   s\, \,{L}^{2}+2  c_0\,   \,D  \, }{ 2\,\left( \sqrt{k\; D}\,L+D\right)  } {e}^{-\sqrt{\frac{\kappa}{D}}\, \left( L-x\right)  }, \ \ x \geq L
	\end{flalign*}
	The maximum concentration is attained at
	\[
	x_m=L - \frac{c_0 \frac{\sqrt{\kappa D}}{s} + \sqrt{\dfrac{\kappa}{D}} \frac{L^2}{2} }{1+ L \, \sqrt{\dfrac{\kappa}{D}}} 
	\]
	It is easy to check that at $x=2 \,x_m$ we have $c_s(2 \, x_m)= c_0$.

\subsection{Fractional diffusion along the line}
\label{sec:frdiff}

The fractional Fick's law in this case reduces to the 
fractional derivative of a function in the sense of Caputo \cite{Caputo1967},\cite{Caputo1971}.
This derivative is in turn defined by the  differ-integral:
\[
\frdiffix{\beta}{ a }  f (x) = \dfrac{1}{\Gamma(1- \beta)}\int_{a}^{x}\frac{  f^{\prime}\left( t\right) }{{\left( x-t\right) }^{\beta}}dt \ecma
\] 
where $\Gamma(.) $ is the Euler's gamma function, which for integer number is equal to the factorial $\Gamma(n) =(n-1)! $

The fractional Fick's law is given in this case by
\[
j= - D \;  \frdiffix{\beta}{ a }  c
\]

For the convenience of the reader a simple derivation is presented in Appendix \ref{sec:fick}.
Combining this equation with a conservation of mass equation for the concentration of particles leads to the fractional diffusion equation of the type.

\[
\frac{\partial c}{ \partial t}  = D  \;  \frac{\partial }{ \partial x}  \frdiffix{\beta}{ 0 } c + s 
\]
In the S compartment the equation yields
\[
\frac{\partial }{ \partial x}  \frdiffix{\beta}{ 0 } c_s +\frac{s}{ D}   = 0 \ecma
\]
with a solution
\[
c_s(x) =\frac{  k_5\ \left(1 + \beta \right)  \,D\,{x}^{\beta}-s\,{x}^{\beta+1}}{ 
	\Gamma \left( \beta +2 \right) \,D}+c_0
\]
For the T compartment
since the first derivative of the equation is constrained at the boundary we will reformulate the problem entirely in terms of Caputo derivatives.
\[
D   \frdiffix{1 + \beta}{ 0 } c_f - \kappa c_f  = 0 \ecma
\]
The equation can be solved in terms of special functions \cite{Mainardi2001,Mainardi1997}.
The general solution is given by
\[
 c_f(x) =  C_1 \, E_{1+\beta, 1} \left(  ^{1+\beta} \sqrt{\frac{\kappa }{D}}\, x |x|^{\beta} \right) + C_2 \, x \, E_{1+\beta, 2} \left(  ^{1+\beta} \sqrt{\frac{\kappa }{D}}\, x |x|^{\beta} \right) 
\]
where the $ E_{a, b}$ denotes the Mittag-Leffler function (see Appendix \ref{sec:LMf} ).
To avoid unbounded solutions at infinity we pick up constants with opposite signs.

Therefore, finally
\[
c_f(x) =  C \left(  E_{1+\beta, 1} \left(  ^{1+\beta} \sqrt{\frac{\kappa }{D}}\, x |x|^{\beta} \right) -  x \, E_{1+\beta, 2} \left(  ^{1+\beta} \sqrt{\frac{\kappa }{D}}\, x |x|^{\beta} \right) \right) 
\]
where the constant \textit{C} can be determined from the boundary or initial conditions. 

Since $\frac{\kappa }{D}$ is assumed to be small we will use further the asymptotic expression valid for small values of $x$ to illustrate the relationship to the integer-order solution. 
\[
 c_f\left( x \right) \approx k_2 e^{-  ^{1+\beta} \sqrt{\frac{\kappa }{D}}\, \Gamma (\beta)  x }
\]

The resulting system can be solved by subsequent integration using eq.  \ref{eq:inverse2} to yield the following algebraic system 
	\begin{flalign*}
	\frac{  k_5\ \left(1 + \beta \right)  \,D\,{L}^{\beta}-s\,{L}^{\beta+1}}{ 
		\Gamma \left( \beta +2 \right) \,D}+c_0 & = k_2\, {{e}^{-\frac{\Gamma\left( \beta\right) \, {{\kappa}^{\frac{1}{\beta+1}}}\, L}{{{D}^{\frac{1}{\beta+1}}}}}} \\
	\frac{  k_5\ \beta  \,D\,{L}^{\beta-1}-  \,s\,{L}^{\beta}}{
		\Gamma \left( \beta + 1 \right) \,D} & = -\frac{ k_2\, \Gamma\left( \beta\right) \, {{\kappa}^{\frac{1}{\beta+1}}}\, {{e}^{-\frac{\Gamma\left( \beta\right) \, {{k}^{\frac{1}{\beta+1}}}\, L}{{{D}^{\frac{1}{\beta+1}}}}}}}{{{D}^{\frac{1}{\beta+1}}}}
	\end{flalign*}
for the first boundary and   second boundary condition, respectively. 
In the resulting system $k_2$ and $k_5$ are unknown constants to be determined by the initial and boundary conditions. 

Considering further  the steady state gives finally for the proximal compartment 
 \[\small
  c_s(x)= \frac{- s\, x^{1+\beta} }{ 
  	\Gamma\left(\beta +2 \right)\,D} 
  +  \frac{  {{L}^{1-\beta}}\, \Gamma\left( \beta\right) \, {{\kappa}^{\frac{1}{\beta+1}}}\, \left( s\, {{L}^{\beta+1}}- c_0\, \Gamma\left( \beta+2\right) \, D\right) +\left( \beta+1\right) \, s\, {{D}^{\frac{1}{\beta+1}}}\, {{L}}   }{\Gamma\left( \beta+2\right) \, D\, \left( \Gamma\left( \beta\right) \, {{\kappa}^{\frac{1}{\beta+1}}}\, L+\beta\, {{D}^{\frac{1}{\beta+1}}}\right) } \, {{x}^{\beta}} +c_0 , \ \ x \in [0, L]
 \]
 while for the distal component
\[
c_f(x) = \frac{\left( s\, {{L}^{\beta+1}}+c_0\, \beta\, \Gamma\left( \beta+2\right) \, D\right) \, }{\Gamma\left( \beta+2\right) \, {{D}^{\frac{\beta}{\beta+1}}}\, \left( \Gamma\left( \beta\right) \, {{\kappa}^{\frac{1}{\beta+1}}}\, L+\beta\, {{D}^{\frac{1}{\beta+1}}}\right) } {{e}^{ - {\Gamma\left( \beta\right) \, {^{\frac{1}{\beta+1}\sqrt{\kappa  \over D}}}\,  }\left( x-L \right)}} , \ \ x \in [ L, \ \infty )
\]

\begin{figure}[h!]
	\begin{center}
	\includegraphics[width=1.0\linewidth, bb=0 0 1533 546]{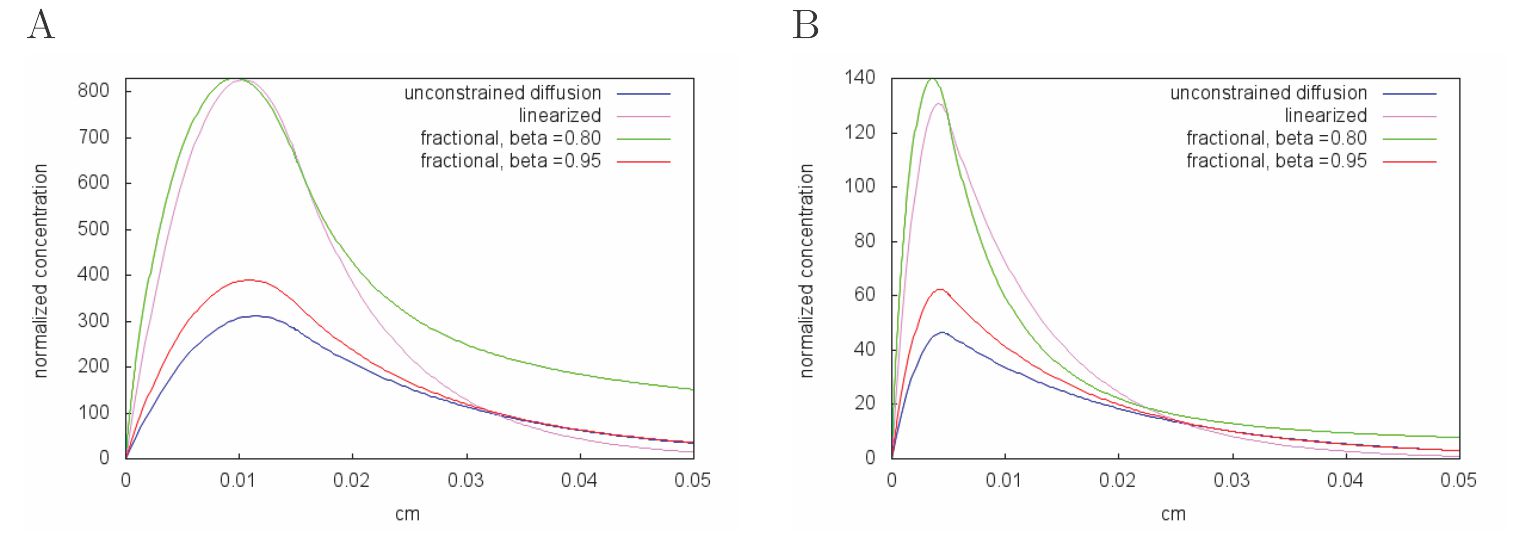}
	\caption[Diffusion models]{Influence of scar thickness on the steady-state concentration of diffusing species in planar geometry}
	\label{fig:diffusion_models1}		 
	\end{center}
The \textit{y-axis} shows normalized concentration of species with regards to source flux intensity $ \frac{c}{s}$; for simplicity also $c_0=0$ was assumed.
Notice the slow decay of the fractional-order solution compared to the linearized case. 
A : L = 150 um, B: L= 50 um.

\end{figure}

\section{Cylindrical geometry}
\label{sec:cylinder}
\subsection{Regular cylindrical diffusion}
\label{sec:cyl}
In the case of implanted wire electrodes, we will assume cylindrical symmetry of the problem.
The the Laplacian operator can be represented as
\[
\nabla^2 c   =  \frac{1}{r} \frac{\partial}{\partial\,r}  \left( r    \frac{\partial \, c}{\partial\,r}     \right)  +\frac{1}{{r}^{2}} \frac{{\partial}^{2} c}{\partial\,{\phi}^{2}}    +\frac{{\partial}^{2} \, c}{\partial\,{z}^{2}}   
\]
%

We will look only for solutions that are cylindrically and axially symmetric that is
$ \dfrac{\partial c}{ \partial \phi} =0 $ and
$ \dfrac{\partial c}{ \partial z} =0 $.


Therefore, we have to consider the following system:
\begin{flalign*}
c (x)  &= c_s(x) + c_f(x) \\
\frac{\partial c_s}{ \partial t} &=	D \frac{1}{r} \frac{\partial }{\partial\,r}  \left( r   \frac{\partial c_s}{\partial \,r}  \right)  + s    \\
\frac{\partial c_f}{ \partial t} &=	D \frac{1}{r} \frac{ \partial}{\partial \,r}  \left( r  \frac{\partial \, c_f  }{ \partial \,r}     \right)   - \kappa \, c_f 
\end{flalign*}

The solution will be presented as a sum of two terms for the two coupled domains where we also impose continuity conditions at the border:
\begin{flalign*} 
\left.  c_s(r) \right|_{r = L-} &= \left.  c_f(r) \right|_{r= L+} \\
\left. \frac{\partial c_s}{ \partial r} (r)\right|_{x = L-} &= \left. \frac{\partial c_f}{ \partial r} (r) \right|_{r = L+}
\end{flalign*}
In addition we assume also a natural boundary conditions in steady state:
\[
c (0) = c_0,  \  \  c(  \infty)  =0
\]
In steady state the equations transform to the following independent system
\begin{flalign*}
D \frac{1}{r}\frac{\partial }{\partial \,r}  \left( r  \frac{\partial \, c_s  }{\partial \,r}    \right) + s =0 , \ & 0 \leq r \leq L   \\
D  \frac{1}{r} \frac{\partial }{\partial \,r} \left( r  \frac{\partial \, c_f  }{\partial \,r}     \right)   - \kappa\, c_f =0 , \ & r \geq L \\
\end{flalign*}

%
\begin{description}

\item [Compartment S]
The general solution is given by 
\[
c_s\left( r \right) =c_1 - \frac{{{r}^{2}} s}{4 D} -   c_2 \ \mathrm{log}\left( r\right) 
\]
From the finiteness of the solution at the left boundary it follows that $c_2=0$ and
\begin{equation}
\label{eq:cyls1}
c_s\left( r \right) =c_0 - \frac{s}{4 D} r^2
\end{equation}

%
\item[Compartment T]  
The equation can be transformed in the form

\[
\frac{\partial}{\partial\,r}  \left( r   \frac{\partial}{\partial\,r} c_f\left( q \,  r\right)   \right) - {{q}^{2}} \, r \,   c_f\left( q \, r\right)=0
\]
with 
\[
q =   \sqrt{ \kappa \over D } ,
\]
which corresponds to the modified Bessel equation \cite{Polyanin2003} by the ansatz
$
 q^2  c_f(q r) =  y (r)
$.
Notably,
\[
r  \, \frac{{{\partial}^{2}\, y}}{\partial\,{{r}^{2}}}   +\frac{\partial \, y}{\partial\,r}  - r \, y  =0
\]

The general solution of this equation is given in terms of the  \textsl{modified Bessel functions} of the first and second kind, notably :
\[
c_f(r) = c_1   K_0 (q r) + c_2   I_0 (q r)
\]
Since $I_0(r)$ diverges at infinity the only acceptable solution is
\[
c_f(r) = c_1   K_0 (q r)
\]

Applying the continuity conditions results in the system
\begin{flalign*}
c_0-\frac{s\, {{L}^{2}}}{4 D} &= c_1 K_0 \left( \sqrt{\frac{ D}{\kappa}}\, L\right)   \\
-\frac{s\, L}{2 D}&=- c_1  \, K_1 \left( \sqrt{\frac{ D}{\kappa}}\, L\right)\,  \sqrt{  \frac{D}{\kappa}  }
\end{flalign*}

The system can be solved to yield:

\begin{flalign*}
c_s(r) & =\frac{s\, (L^2 -r^2)}{4 D}+\frac{K_0\left( \sqrt{  \kappa \over D } \, L\right) \, s\, L}{ K_1\left( \sqrt{  \kappa \over D } \, L\right) \, \sqrt{\kappa D}  } , \ r \in [0, L]  \\
c_f(r) & = \frac{ K_0\left( \sqrt{  \kappa \over D } \, r\right) \, s\, L}{ K_1\left( \sqrt{  \kappa \over D } \, L\right) \, \sqrt{ D \kappa} } , \ r > L
\end{flalign*}

\end{description}
From the explicit form of solution it can be seen that the value of the concentration at the origin is completely fixed by the geometry of the problem.

\subsection{Fractional cylindrical diffusion}
\label{sec:frdiffcyl}

Generalization to the 2D case is a considerably more difficult problem.
Following the derivation of Meerschaert \cite{Meerschaert2006} we define the fractional Fick's law of fractional order $\beta <1$ as
\[
\nabla^{ \beta}  = \mathbb{J}^{1 -\beta}   \nabla 
\]
where $\mathbb{J}^{\beta} $ denotes the vector fractional integral operator. 
\[
\frac{\partial c}{ \partial t}  = D  \;  \nabla \cdot \nabla^\beta  c + s- \kappa c 
\]

\begin{description}
	\item[Compartment S] 
	Following the derivation of Meerschaert \cite{Meerschaert2006} we define the fractional Fick's law.
	Therefore, in steady state we have 
	\[
	D  \;  \nabla \cdot \nabla^\beta  c_s + s  =0
	\]
	The equation can be solved by Laplace's transform method to yield
	
	\[
	c_s (r) = c_0 -\frac{{{r}^{\beta+1}}\, s}{ D \,(1 + \beta)\, \Gamma \left( \beta+2\right) } + \frac{ c_1 \,{{r}^{\beta-1}}\, s}{ D \,(1 - \beta)\, \Gamma \left( \beta\right) }
	\]
	Similar considerations about the boundedness of the solution at $r=0$ require that $c_1=0$.
	\begin{equation}
	\label{eq:fractionborder}
	c_s (r) = c_0 -\frac{{{r}^{\beta+1}}\, s}{ D \,(1 + \beta)\, \Gamma \left( \beta+2\right) } 
	\end{equation}
	
	\item[Compartment T]  
	In steady state conditions we have
	\[
	D  \;  \nabla \cdot \nabla^\beta  c_f -  \kappa \, c_f  = 0
	\]
	
The Laplace transform method can be also applied in this case, although no explicit solution can be identified (
see Appendix \ref{sec:fractional-order-case}). 
Nevertheless, for completeness of the presentation a similar procedure is applied.
This is equivalent to the assumption that the fractional exponent equals two in the T domain. 

Applying the continuity conditions results in
\begin{flalign*}
c_0 -\frac{{{L}^{\beta+1}}\, s}{(1+\beta)\, D \,\Gamma \left( \beta+2\right) }  &= c_1 K_0 \left( \sqrt{\frac{ D}{\kappa}}\, L\right)   \\
-\frac{s\, L^\beta}{(1+\beta) D \Gamma(\beta+1)}&=- c_1    K_1 \left( \sqrt{\frac{ D}{\kappa}}\, L\right)\,  \sqrt{ \frac{D}{\kappa} }
\end{flalign*}  
	
\end{description}

 \begin{figure}[h!]
 	\begin{center}
 \includegraphics[width=1.0\linewidth, bb=0 0 1533 546]{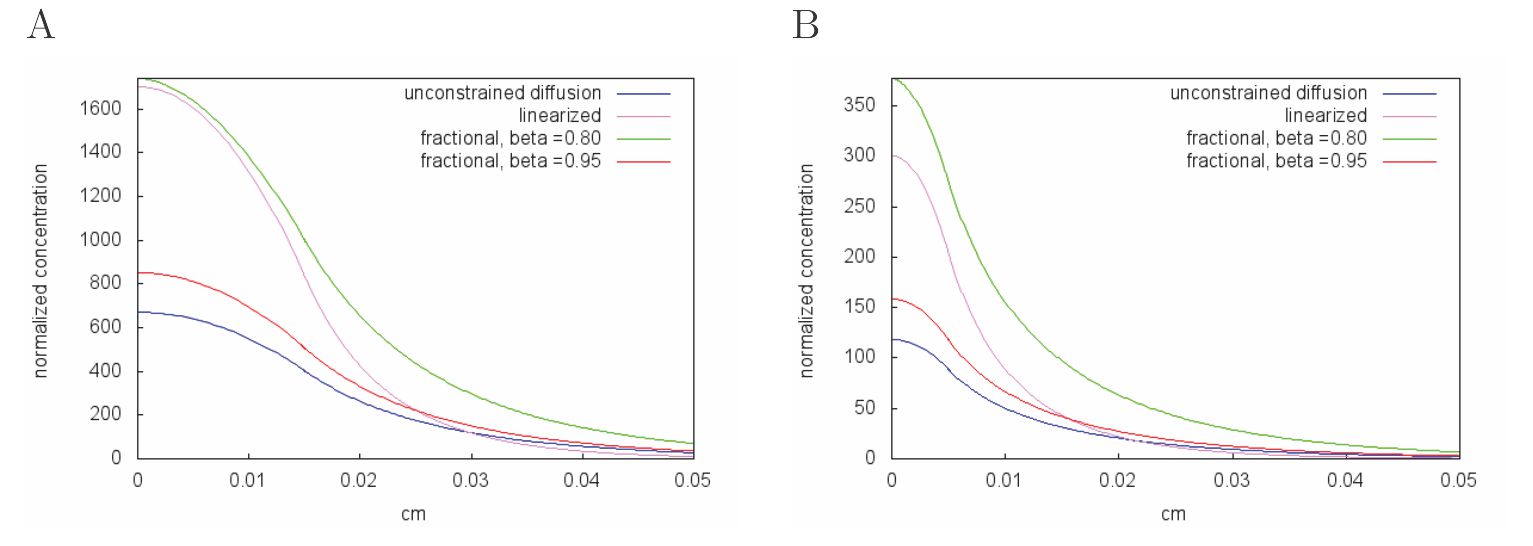}
 	\caption[Diffusion models]{Influence of scar thickness on the steady-state distribution of diffusing species in cylindrical geometry}
 	\label{fig:diffusion_models2}
	 \end{center}
	The \textit{y-axis} shows normalized concentration of species with regards to source flux intensity $ {s}$.
	A : L = 150 um, B: L= 50 um.
 \end{figure}
 
 \section{Analysis of the boundary conditions in the case of a thick electrode}
 \label{sec:assumption1}
 
 %
 \subsection{Vanishing flow}
 
 Let's assume that the electrode is thick, with a finite radius $\rho$.
This case is known as the standard Neumann boundary condition
\[
\left. \frac{ \partial \, c_s (r) }{ \partial r}\right|_{r=\rho}=0 
\]

 \begin{description}
 	\item[Linear geometry] 
 In the case of tight contact between the tissue and the electrode one can assume as well vanishing of the flux at the boundary of the electrode. 
 In this case we will have 
 
 \[\frac{\sqrt{k}\, \left( s\, {{L}^{2}}-2\, c_0\, D\right) +2\, s\, \sqrt{D}\, L}{2\, \sqrt{k}\, D\, L+2\, {{D}^{\frac{3}{2}}}}-\frac{\rho\, s}{D} =0 \]
 
 which gives immediately for the concentration at the boundary
 
 \[c_s (\rho)=\frac{s\, \left( L-\rho\right) \, \left( \sqrt{k}\, L+2\, \sqrt{D}-\sqrt{k}\, \rho\right) }{2\, \sqrt{k}\, D}\]
 
 resulting in
 \[
 c_s (x) = 
 \frac{s\, \left( L-x\right) \, \left( L+x-2\, \rho\right) }{2\, D}+\frac{s\, \left( L-\rho\right) }{\sqrt{k}\, \sqrt{D}}
 \]
 The maximum of the last expression is attained at $x=\rho$. 
 
 \item[Cylindrical geometry] 
 
  The general solution for the S-compartment is given by 
  \[
  c_s\left( r \right) =c_1 - \frac{{{r}^{2}} s}{4 D} -   c_2 \ \mathrm{log}\left( r\right) 
  \]
 
  After some algebraic manipulations the solution can be simplified to
  
  \[ 
  c_s (r) = 
  \frac{\mathrm{log}\left( \frac{{{r}^{2}}}{{{\rho}^{2}}}\right) \, {{\rho}^{2}}\, s}{4\, D}+\frac{ \left( {{\rho}^{2} - r^2} \right) \, s}{4\, D} +c_0
  \]
  It can be shown that in the limit $\rho \rightarrow 0$ we can recover
  \[ 
  c_s\left( r \right) =c_0 - \frac{s}{4 D} r^2
  \]
  
  In the case of fractional diffusion after some algebra we can obtain
  
  \[c_s(r) =c_0 
  - \frac{  2\, {{\rho}^{\beta+1}}\, s}{\left( {{\beta}^{2}}-1\right) \, \Gamma\left( \beta+2\right) \, D} 
  +\frac{{{r}^{\beta-1}}\, {{\rho}^{2}}\, s}{\left( \beta-1\right) \, \Gamma\left( \beta+2\right) \, D}
  -\frac{{{r}^{\beta+1}}\, s}{\left( \beta+1\right) \, \Gamma\left( \beta+2\right) \, D}
  \]
  which again for $\rho \rightarrow 0$ and $\beta \rightarrow 1$ reduces to Eq. \ref{eq:cyls1}.
 \end{description}

 %
 \subsection{Non-vanishing flow}
 
 Let's denote  by $z$  the value of the flow attained at the electrode boundary.
 In this case we have to consider the condition
 \[
 \left. \frac{ \partial \, c_s (r) }{ \partial r}\right|_{r=\rho}= z
 \]
 
 \begin{description}
 	\item[ Linear geometry]  
 	 In this case we have the condition
 	 \[
 	 \frac{\sqrt{k}\, \left( s\, {{L}^{2}}-2\, k_3\, D\right) +2\, s\, \sqrt{D}\, L}{2\, \sqrt{k}\, D\, L+2\, {{D}^{\frac{3}{2}}}}-\frac{\rho\, s}{D}=z
 	 \]
 	 After some algebra we arrive at:
 	 \[ 
 	 c_s(x) =\sqrt{k}\, {{D}^{\frac{3}{2}}}\, \left( s\, L-z\, D-\rho\, s\right) +\frac{\left( L-x\right) \, \left( s\, L-2\, z\, D+s\, x-2\, \rho\, s\right) }{2\, D}
 	 \]
 	 
 	 The position of the peak is attained by solving the equation
 	 \[
 	 \frac{s\, \left( L-x\right) }{2\, D}-\frac{s\, L-2\, z\, D+s\, x-2\, \rho\, s}{2\, D} = 0
 	 \]
 	 which gives the unique value
 	 \[
 	 x_m=\frac{z\, D}{s}+\rho
 	 \]
 	 \item[Cylindrical geometry ]  
 	In this case we have the condition
 	\[-\frac{\rho\, s}{2\, D}-\frac{c_2}{\rho}=z\]
 	and after some algebraic manipulations  
 	\[ c_s(r) =\frac{\left( \left( \mathrm{log}\left( \frac{{{r}^{2}}}{{{\rho}^{2}}}\right) +1\right) \, {{\rho}^{2}}-{{r}^{2}}\right) \, s}{4\, D}+ \mathrm{log}\left( \frac{{{r} }}{{{\rho} }}\right) \, \rho\, z  + c_0\]
 	It can be shown that in limit $\rho \rightarrow 0$ we can recover
 	\[ 
 	c_s\left( r \right) =c_0 - \frac{s}{4 D} r^2
 	\]
 	
 	The position of the peak is attained by solving the equation
 	\[
 	\frac{2\, \rho\, z\, D+{{\rho}^{2}}\, s}{2\, r\, D}-\frac{r\, s}{2\, D} = 0
 	\]
 	which gives the physically meaningful value of
 	\[
 	r_m =\sqrt{\rho} \sqrt{\frac{2\, z\, D}{s}+{{\rho}}}
 	\]
 	
 \end{description}
 From this analysis it is apparent that the peak position is influenced by  the direction of the flux.
 
 \section{Numerical analysis}
 \label{sec:numer1}
 
 Some numerical results are demonstrated in Fig. \ref{fig:diffusion_models1}  using literature values of parameters for rat neocortex, \cite{Nicholson2001}.
 
 Available experimental data show that a typical value for small molecules for $\lambda$ in the nervous system is about 1.6, which implies that $\bar{D}$ is some 2.6 times smaller than \textit{D}. For macromolecules $\lambda$ is increased, in part because of more frequent interaction with the walls of the narrow channels through the ECS
 \cite{Nicholson2011}.  
 Numerical values used in the calculation are as follows:  $D=1. 10^{-6} cm^2 s^{-1}$ for proteins, $\lambda = 2.5$ (NGF) and $\alpha=0.21$; uptake $k=1. 10^{-4}$ (NGF). 
 Magin et al. \cite{Magin2013} measure experimentally the fractional order of 1.95 using diffusion MRI.
 This corresponds to the value $\beta=0.95$ in our parametrization.
 
 All cases use the same values of porosity. The glial scar thickness  was taken as the range $L= 50 - 150 \; \mu m$.
 For the unconstrained diffusion case no correction for $\lambda$ is given.
 The linearized case corresponds to the approach of Nicholson and Sykova \cite{Nicholson2001, Sykova2008}.
 In the fractional cases no tortuosity correction is applied but only the fractional order $\beta$ is varied.
 Results are plotted in Figs. \ref{fig:diffusion_models1} and \ref{fig:diffusion_models2}.
 
\section{Discussion}
\label{sec:disc}

%
\subsection{Diffusion around brain implants}
\label{sec:disimplant}
 
Presented results show that in steady-state   conditions substances, which are produced presumably in the glial scar could accumulate and produce characteristic profiles, matching qualitatively empirically observed 
GFAP and microglial profiles published in available experimental literature.
More specifically, shapes of the concentration profiles correspond qualitatively with the GFAP profiles, for example \cite{Zhong2007, McConnell2009a, Winslow2010, Winslow2010a, Potter2012,Potter2013,Skousen2015}, and the blood vessel area distribution \cite{Grand2010} reported in literature.
In all of these studies implanted probes had planar geometry, therefore reported GFAP profiles can be discussed under the assumptions of the presented model.
Presented model can be extrapolated to separable 3D geometries as well considering presence of planar symmetries. 
The shape of the cylindrical distribution corresponds also to studies employing cylindric or wire electrodes \cite{Rao2012,Kim2004}. 
Presented approach, therefore, provides means for approximation of such profiles and estimation of geometrical parameters related to the problem, such as the scar thickness and the intensity of the source.
In some cases the fractional exponent can supplement the so-far used linearized parameters, such as tortuosity, if such estimates are not available.

Our model demonstrates that ex-centric accumulation of diffusing species can occur and the scar geometry (i.e. tortuosity,  fractional order, scar thickness) can influence such accumulation.
From the analysis of the boundary condition we can conclude that the insulation of the electrode with regard to the boundary (i.e. presence or absence of lateral flux) determines the shape of the observed profile in both studied geometries. 
We can therefore speculate, that all studies exhibiting profiles with ex-centric accumulation actually had somehow loose contact between the implant and the brain tissue, possibly due to micromotion.
To further verify that, however, additional efforts in mechanical modeling of devices are necessary.

Recently \cite{Skousen2015} modeled numerically diffusion around implants in order to design a diffusion sink placed at the device surface that would retain pro-inflammatory cytokines for sufficient time to passively antagonize their impact on the foreign body response. 
Such setting would correspond to a negative flux condition in our model and it can indeed be shown that this will result in reduction of the gradient. 
Authors used no-flux boundary conditions and presented results, which qualitatively correspond to the analysis presented here. On the other hand, reported experimental results demonstrate deviations from the numerical model, which can be interpreted well using our model.

\subsection{Impeded diffusion phenomena} 
 \label{sec:imdisc}
 
 Anomalous diffusion problems naturally arise in the setting of complex biological environment. 
 Modeling of diffusion in different complex media could provide further understanding in a variety of experimental conditions.
 Impeded or anomalous diffusion in the brain is already a well established phenomenon \cite{Sykova2008}.  
 Sykova and Nicholson \cite{Sykova2008} list several factors that determine diffusion impediments:   
 i) an increase in geometric path length; ii) transient trapping of molecules in dead-space microdomains; iii) an increased interstitial viscous drag on migrating molecules; iv) transient binding to membrane-attached or extracellular matrix-attached receptors; and v) nonspecific interaction with fixed opposite charges. 
 All five factors can be thought of as introducing a delay into the passage of a molecule in brain tissue relative to that in a free medium.
 Therefore, this complex medium can be conceptually modeled as the presence of two phases -- one permeable and one impermeable, which impedes undergoing diffusion.
 
 Fractional diffusion models have been employed in hydrology describing well slow diffusion \cite{Meerschaert2006} in protein diffusion in the plasma membrane \cite{Kou2004,Khoshnood2013}.
 The spatial complexity of a medium can impose geometrical constraints on transport processes on all length scales that can fundamentally alter the laws of standard diffusion \cite{Metzler2004}. 
 Specifically fractional order diffusion models have been employed to model water diffusion in the brain \cite{Magin2008}.
  
 Fractional models can provide new insights on mesoscopic aspects of the studied phenomena. For example, space fractional models have been used to describe cardiac cell conduction \cite{Bueno-Orovio2014}, and diffusion phenomena in fixed tissue samples \cite{Magin2013}. 
  
 ECS of the brain comprises the matrix that resides outside the neurons and glia cells, most importantly the interstitial space between neighboring cells. 
 The ECS is a reservoir for ions involved in electrical activity, a communication channel for chemical messengers and a conduit for drug delivery.
 A quantitative description of extracellular diffusion is important whenever the transport of neurotransmitters, neuromodulators, and therapeutics in the brain  ECS is considered.
 Diffusion-mediated transport of biomolecules substances is hindered by the ECS structure but the microscopic basis of this hindrance is not fully understood \cite{Hrabetova2003}. 
 Evidence for anomalous diffusion in the brain has been found in the rat cerebellum \cite{Xiao2015}, which provides support also for employing fractional models in the continuous approximation limit.
 Most probably, dead space microdomains can be the cause of such anomalous diffusion \cite{Sherpa2014}.
 
 The magnitude of the exponent $\beta$ can be used as a metrics for the departure from linearity. 
 Since decrease of $\beta$ will result in behavior approximating advection flow, this can used to reason about state where such flows can be enhance, for example in the case of the leaky blood-brain barrier or in conditions with increased local blood flow.
 
  \section*{Acknowledgments}
  The work has been supported in part by a grant from Research Fund - Flanders (FWO), contract number 0880.212.840.
    
\appendix

\section{Derivation of the fractional-order Fick law}
\label{sec:fick}

The Riemann-Liouville fractional integral (or differ-integral)  defines a weighted average  of the function over the interval \textit{[a, x]} using a power law  weighting function.

The Riemann-Liouville differintegral of order $\beta \geq 0$ is defined \cite{Oldham1974} as
\[
\frdiffiix{\beta}{ a }  f (x) = 
\dfrac{1}{\Gamma(\beta)} \int_{a}^{x}   f \left( t \right)  \left( x-t \right)^{\beta -1}dt \epnt
\] 

The non-local fractional derivative of a function in the sense of Riemann-Liouville is defined in terms of the fractional integral as 
\begin{equation}
	\frdiffixr{n +\beta}{ a }   f (x) = \left( \frac{d}{dx} \right)^n \frdiffiix{n-\beta}{ a }  f (x)
\end{equation}
which is usually specialized for $n=1$   in an explicit form by
\[
\frdiffixr{\beta}{ a }  f (x) = \dfrac{1}{\Gamma(1- \beta)}  \frac{d}{dx}  \int_{a}^{x}\frac{  f \left( t\right) }{{\left( x-t\right) }^{\beta }}dt \epnt
\]

The non-local fractional derivative of a function (in the sense of Caputo)  of a function is defined   \cite{Caputo1967},\cite{Caputo1971} as
\begin{equation}
	\frdiffix{\beta}{ a }  f (x) =  \frdiffiix{n-\beta}{ a }  f^{(n)} (x) \epnt
\end{equation}
or using expanded notation for $n=1$
\[
\frdiffix{\beta}{ a }  f (x) = \dfrac{1}{\Gamma(1- \beta)}\int_{a}^{x}\frac{  f^{\prime}\left( t\right) }{{\left( x-t\right) }^{\beta}}dt \ecma
\] 
where $\Gamma(.) $ is the Euler's function, which for integer number is equal to the factorial $\Gamma(n) =(n-1)! $. 

Both definitions coincide for problems where the function and its first $n$ derivatives vanish at the lower limit of integration, i.e. when $f(a)=0, \ldots , f^{(n)} (a) =0 $.
Caputo's definition is better suited for problems where the function and its derivatives doe not vanish at this limit, because in this case it is given as kind of regularization of the Riemann-Liouville differintegral  to avoid divergences.

Caputo's derivative is a left inverse of the fractional integral:
\begin{equation}
	\label{eq:inverse1}
	\frdiffix{\beta}{ a } \circ \frdiffiix{\beta}{ a }    f = f(x) \ecma
\end{equation}
while the  fractional integral is a conditional inverse of Caputo's derivative:  
\begin{equation}
	\label{eq:inverse2}
	\frdiffiix{\beta}{ a } \circ  \frdiffix{\beta}{ a }  f = f(x) - f(a^+)
\end{equation}
in the case when $\frdiffix{\beta}{ a }  f \neq 0 $. 
This allows one to solve simple fractional systems, such as the fractional diffusion equation in the example. 

Fractional Fick's law for an exponent $\beta>0$ can be derived under the assumption that the transport is given by a wighted fractional average of the differential of concentrations in the domain $[a, x]$.
To avoid unphysical divergence at the borders of the domain, which are of interest in our problem we further regularize by subtracting the boundary concentration $c(a)$:
\[
dJ_\beta = dx \dfrac{1}{\Gamma(\beta)} \frac{d}{dx}  \int_{a}^{x}  \left(  c \left( t \right) - c\left( a \right) \right)  \left( x-t \right)^{\beta -1}dt \epnt
\]
One can recognize this expression as the usual definition of the Riemann-Liouville fractional derivative.
The integral can be evaluated partially assuming as usual existence of the derivative of the concentration $c^\prime(t)$.
Then applying integration by parts we get
\begin{flalign*}
	\dfrac{1}{\Gamma(1 + \beta)}   \int_{a}^{x}  \left(  c \left( t \right) - c\left( a \right) \right)  d (x-t)^{\beta}   & = \\
	\left.  \dfrac{1}{\Gamma(1 + \beta)}  \left(  c \left( t \right) - c\left( a \right) \right) (x-t)^{\beta}  \right|^{x}_{a}  - 
	\dfrac{1}{\Gamma(1 + \beta)}  \int_{a}^{x} c^\prime \left( t \right) (x-t)^{\beta} dt = \\
	- \dfrac{1}{\Gamma(1 + \beta)}  \int_{a}^{x} c^\prime \left( t \right) (x-t)^{\beta} dt
\end{flalign*}
where we notice that the first term evaluates to zero.

Differentiating the last integral by $x$ gives
\begin{flalign*}
	dJ_\beta = - dx \dfrac{1}{\Gamma(1+ \beta)} \frac{d}{dx}   \int_{a}^{x} c^\prime \left( t \right) (x-t)^{\beta} dt = \\
	- dx \dfrac{\beta}{\Gamma(1+ \beta)} \int_{a}^{x} c^\prime \left( t \right) (x-t)^{\beta-1} dt = 
	- dx \dfrac{1}{\Gamma( \beta)} \int_{a}^{x} c^\prime \left( t \right) (x-t)^{\beta-1} \epnt
\end{flalign*}

However, the last expression can be recognized as $\frdiffix{1- \beta}{ a }  c$.
Therefore,
\begin{equation}
	\label{eq:frfick}
	j = \dfrac{dJ_\beta}{dx} = - \frdiffix{1- \beta}{ a }  c
\end{equation}

\section{Mittag-Leffler functions}
\label{sec:LMf}

During the last decades fractional calculus has been widely applied in many scientific areas ranging from mathematics and physics, up to biology, engineering, and earth sciences. 
The Mittag-Leffler functions play an important role in fractional calculus since many solutions of fractional differ-integrals can be expressed in terms of Mittag-Leffler functions.

The one parameter ML function is a generalization of the exponential function:
\begin{equation}
	\label{eq:ml1}
	E_{\alpha} (t) = \sum\limits_{k=0}^{\infty} \frac{t^k}{\Gamma (\alpha k +1 )}, \alpha >0
\end{equation}

The two parameter ML function is an additional generalization in the sense
\begin{equation}
	\label{eq:ml2}
	E_{\alpha, \beta} (t) = \sum\limits_{k=0}^{\infty} \frac{t^k}{\Gamma (\alpha k + \beta )}, \alpha >0
\end{equation}

A particular form of the one parameter ML function studied by F. Mainardi is
\[
e_\alpha (t) =  E_{\alpha} (- t^ \alpha)
\]

The following result can be stated 
\[
\frdiffix{ \alpha}{ 0 }  e_\alpha ( b \; t)  = b \; e_\alpha ( b \; t).
\]

It is common to point out that the function eα(t) matches for $t \approx 0$ a stretched  
whereas as $t \rightarrow \infty$ with a negative power law. 
The short time approximation is derived from the power series representation as follows \cite{Mainardi1997}
\[
e_\alpha (t) \sim \exp \left(  - \frac{t^\alpha}{\Gamma( 1 + \alpha)} \right) , \ \ t \rightarrow 0
\]
\[
e_\alpha (t) \sim  \frac{1}{1 + \Gamma( 1 - \alpha) \, t^\alpha}  , \ \ t \rightarrow \infty
\]

For intermediate ranges, the following approximations can give acceptable results for $\alpha \approx 1$
\[
e_\alpha (t) \approx \dfrac{{e^{-\Gamma\left(1 + \alpha\right) \, {t^{\alpha}}}}+
	\frac{\Gamma\left(1 + \alpha\right) }{ 1 + \Gamma\left( 1-\alpha\right) \, {t^{\alpha}}}}{
	1 + \frac{\Gamma\left( 1+ \alpha \right) }
	{1 +\Gamma\left( 1-\alpha\right) \, {{t}^{\alpha}}}}
\]
and
\[
e_\alpha (t) \approx \frac{e^{  - \Gamma\left( a\right) \, x   }+\frac{1}{1+\frac{x}{a}}}{1+\frac{1}{1+\frac{x}{a}}}
\]

%
\section{Laplace transform method for solving fractional differential equations}
\label{sec:laplace}

Since we are dealing with a problem in the positive half-plane the Laplace integral transform method can be applied for solving all presented differential equations.
We further outline the main steps of the solution procedures.

\subsection{Integer order case}\label{sec:integer-order-case}
Considering the flux conditions the following equation can be stated for the Green's function:
\[   
\frac{  D }{r} \frac{\partial \, f}{\partial\,r}  + D\, \frac{{{\partial}^{2} f }}{\partial\,{{r}^{2}}} +s = - c_1 \frac{\delta\left( r\right) }{r}
\]
the Laplace transform of the equation is as follows 	
\[- D \, {{p}^{2}}   \frac{\partial \, F}{\partial\,p} - D\, p\, F+\frac{s}{{{p}^{2}}}=-c_1\]
which can be solved to give

\[F(p) =-\frac{s}{2\, {{p}^{3}} \, D}+ c_1\,\frac{ \mathrm{log}\left( p\right) }{p\, D}+\frac{c}{p}
\]
the inverse Laplace transform gives

\[ f(r) = -\frac{{{r}^{2}}\, s}{4\, D}-\frac{c_1 \, \mathrm{log}\left( r\right) }{D}+ c\]

\subsection{Fractional order case}\label{sec:fractional-order-case}
The fractional diffusion equation can be transformed in the Laplace domain using 
the properties of the Laplace transform for the Caputo fractional derivative.
Briefly, if $ \mathcal{L} : f(t) \mapsto F (s) $ denotes the Laplace transform then 
the Caputo derivative is transformed according to the following rule:
\[
\mathcal{L} :  \frdiffix{n+ \beta}{0 } f (x)  \mapsto
{{s}^{\beta -1}}\, \left({{s}^{n+1}}\, F (s) - \sum_{k=0}^{n} \mathrm{f}^{(k)}\left( 0 \right) \, s^{n-k}   \right)
\]

\begin{description}
	\item[S--compartment]  
	The Green's function equation in the spatial domain reads
	\[
	\frac{D}{r} \dfrac{\partial }{ \partial r } r \frdiffix{\beta}{a } f (r) + s = - c_1 \frac{\delta(r)}{r}
	\]
	The Laplace transform gives the ordinary differential equation
	\[
	- D \, {{s}^{\beta+1}}\,  \frac{\partial \, F}{\partial\,p}   - D \, {{p}^{\beta}}\, F+\frac{s}{{{p}^{2}}}=-c_1
	\]
	
	with a solution
	
	\[
	F(p) =\frac{c_0}{p} -\frac{s \, {{p}^{-\beta-2}}}{D \left( \beta+1 \right) } -\frac{c_1 \, s}{D \left( \beta-1\right) \, {{p}^{\beta}}}
	\] 
	
	Which gives
	\[
	f(r) = c_0 -\frac{ s\, {{r}^{\beta+1}}}{D \left( \beta+1\right) \, \Gamma\left( \beta+2\right) } 
	- \frac{ c_1 s \,{{r}^{\beta-1}}}{{{ D \left( 1 - \beta\right) } }\, \Gamma\left( \beta    \right) }
	\]
	
	\item[T--compartment]

	The Fractional diffusion equation for the Green's function  reads
	\[
	\frac{D}{r} \dfrac{\partial }{ \partial r } r \frdiffix{\beta}{a } f (r) - \kappa f(r) = - c_2 \frac{\delta(r)}{r}
	\]
	which can be expanded into
	\[
	r D  \frdiffix{1 + \beta}{a } f (r)  +  D \frdiffix{ \beta}{a } f (r) - \kappa r  f(r) = - c_2 \delta(r)
	\]
	This equation in turn can be transformed into the Laplace domain as 
	\[
	\left( \kappa - D \, {{ p }^{\beta+1}} \right) \frac{\partial\, F}{\partial\,p} - D \, {{p}^{\beta}}\, F= -c_2
	\]	
	The last equation can be solved in special functions, notably
	\begin{flalign*}
		F(p) = c_2 \frac{1}{{{\left( {D \, {p}^{1 +\beta}}- \kappa \right) }^{\frac{1}{\beta+1}}}} + \\
		c_1 \frac{p}{k ^{\beta/(1+\beta)} {{\left( { D \, {p}^{1 +\beta}}-\kappa \right) }^{\frac{1}{\beta+1}}}} \
		{_{2}{F}_{1}}\left( \frac{1}{\beta+1},\frac{\beta}{\beta+1}; \frac{1}{\beta+1}+1;{ \frac{D}{\kappa} \, {p}^{\beta+1}}\right) 
	\end{flalign*}
	where $ _{2}{F}_{1} (a,b; c; z) $ is the Gauss hypergeometric function \cite[ch. 15]{Olver2010}.
	Unfortunately, the inverse Laplace transform of the general solution is not known. 
	However, as a verification of so-obtained solution for the case $\beta=D=\kappa = 1$ we can obtain \cite[ch. 15]{Olver2010}:
	\[
	F(p) = c_2 \frac{1}{\sqrt{p^2-1}} +c_1 \frac{\mathrm{acosh}(p)}{\sqrt{p^2-1}} \epnt
	\]
	
	Transforming back to the spatial domain this gives a general solution in terms of modified Bessel functions:
	\[
	f(r) = c_2 I_0 (r) + c_1 K_0 (r) \epnt
	\]
	
\end{description}

 \bibliographystyle{plain}

\end{document}